\documentclass{PoS}
\usepackage{atlasphysics}

\title{Early Search for Supersymmetry at ATLAS}

\ShortTitle{Early Search for Supersymmetry at ATLAS}

\author{\speaker{Xuai Zhuang ~~on behalf of the ATLAS Collaboration}
  \thanks{ The author would like to thank the organizing committee of the 
    Workshop on Discovery Physics at the LHC -Kruger 2010 
    and everyone whose work contributed to this paper.}\\
       Ludwig-Maximilians-Universit\"{a}t, M\"{u}nchen, Germany\\
        E-mail: \email{Xuai.Zhuang@physik.uni-muenchen.de}}

\abstract{
  The search for physics beyond the Standard Model (BSM) is one of the 
most important goals for the general purpose detector 
ATLAS at the Large Hadron Collider at CERN. Supersymmetry search strategies 
based on generic event signatures of high jet multiplicity 
and large missing transverse momentum, optionally including 
leptons in the final state with R-parity conservation are discussed 
in this document. We review the results for above SUSY search 
strategies with first data up to 305 $nb^{-1}$ of integrated luminosity collected by ATLAS 
during 2010 at a centre-of-mass energy $\sqrt{s}$ = 7 TeV.
}

\FullConference{Workshop on Discovery Physics at the LHC -Kruger 2010\\
		December 05-10, 2010\\
		Kruger National Park, Mpumalanga, South Africa}

\begin{document}

\section{Introduction}
Supersymmetry (SUSY) 
is a theoretically favoured candidate for physics 
beyond the Standard Model. If strongly interacting supersymmetric particles 
are present at the TeV-scale, then such particles should be copiously produced 
in the 7 TeV collisions at the Large Hadron Collider (LHC) at CERN \cite{LHC}. 
Therefore the search for the Supersymmetric particles is one of the most important aims 
of the ATLAS experiment at LHC. 

This document presents a first comparison of data to Monte Carlo simulations 
for some of the most important kinematical variables that are expected 
to be employed in supersymmetry searches involving jets and missing transverse 
momentum and possibly including isolated leptons (electrons or muons)
\cite{SUSY0L}\cite{SUSY1L}\cite{SUSYBJET}. Selections based
on these variables are expected to be sensitive not only to R-parity conserving 
SUSY particle production, but also to any model in which one or more 
strongly-interacting particles decay semi-invisibly producing leptons and jets.
Three distinct signatures will be discussed in this document, which
differ by the number of leptons in the final states (no-lepton
and one-lepton channels) and by the application of a further b-tagging
requirement (b-jet channel). 
The measurements in this document are based on data collected in the 
proton-proton collisions at $\sqrt{s}$ = 7 TeV at the LHC 
before August 2010. The no-lepton and one-lepton channels 
used a total integrated luminosity of $70\pm8$ $nb^{-1}$ while the
b-jet channel used $305\pm17$ $nb^{-1}$.
An evaluation of the discovery potential of supersymmetry for the channels
described above with 1 $fb^{-1}$ of a total integrated luminosity  
with the ATLAS detector is also presented\cite{SUSYPros}.
\section{Inclusive SUSY search strategy and first measurements}
\subsection{Inclusive SUSY search strategy}
High-energy jets, missing transverse energy (\met) and possibly leptons 
are the typical signature of R-parity conserved SUSY events at the LHC. 
The observation of deviations from the Standard Model may manifest 
the presence of SUSY. The discovery of new physics can only be claimed 
when Standard Model backgrounds are understood well and under control.
Some of the most important SUSY sensitive variables have been measured 
and compared between data and Monte Carlo simulations to understand the 
Standard Model backgrounds.
Due to the limited statistics of data taking considered here, the prediction 
of the Standard Model backgrounds is derived from Monte Carlo simulation directly.

One of the important SUSY sensitive variables is missing transverse momentum (\met), 
which is formed from two components. The first component is obtained
from the vector sum of the transverse energies of all three-dimensional 
topological clusters in the calorimeter. 
The second component is obtained from the vector sum of the transverse momenta of isolated  
muons. The lepton isolation requirements is discussed later.
The total missing transverse momentum is
computed by a vector sum of these two components. Another important SUSY sensitive 
variable is the effective mass 
($\meff=\sum{^{N_{jets}}_{i=1}p^{jet,i}_{T}}+\sum{^{N_{leps}}_{j=1}p^{lep,j}_{T}}+\met$), 
which is the scalar sum of the transverse momenta 
of $N_{jets}$ leading jets, $N_{leps}$ leading leptons and \met, 
where $N_{jets}$ and $N_{leps}$ are the number of jets and the number of 
leptons required in the analysis seperately. The transverse mass of the lepton and the \met~
($\mt\ =\sqrt{2\cdot p_{\mathrm{T}}^\ell\cdot\met\cdot(1 - \cos(\Delta\phi(\ell,\met)))}$) 
is also used to suppress Standard Model backgrounds in the one-lepton channel.
\subsection{Event selection and systematic uncertainty}
After the good objects selection, the corresponding trigger requirement and a set of 
cleaning cuts to reject events containing jets which are consistent with 
calorimeter noise, cosmic rays or out-of-time energy deposits, the events 
are preselected by asking for at least two jets with transverse 
momentum \pt~ $>$ 30 GeV and one isolated lepton (electron or muon) with
\pt~ $>$ 20 GeV in the one-lepton channel and in the b-jet channel, and by
asking for at least two (2 jet channel), three
(3 jet channel) or four (4 jet channel) jets with \pt~ $>$ 30 GeV,
leading jet \pt~ $>$ 70 GeV and no lepton with \pt~ $>$ 10 GeV in the
no-lepton channel.
The signal region of the one-lepton channel is then defined by applying 
two further cuts: \met~ $>$ 30 GeV and \mt~ $>$ 100 GeV. For the b-jet channel 
with one lepton, another two cuts are requiring: at least one b-tag 
with the decay length significance $SV_0$ ($= L/\sigma{(L)}$, L is the integrated luminosity) 
>6 and \met~ significance $\met/\sqrt{\sum{E_T}} > 2 \sqrt{GeV}$.
The signal region of the no-lepton channel is defined by requiring: \met~ $>$ 40 GeV, 
$\Delta\phi(\textrm{jet}_{i},\vec{E}_{\mathrm{T}}^{\mathrm{miss}}) > 0.2 \, (i=1,2,3)$ 
and $\met/\meff>0.3$ (2 jet channel), 0.25 (3 jet channel) or 0.2 (4 jet channel), 
where $\Delta\phi(\textrm{jet}_{i},\vec{E}_{\mathrm{T}}^{\mathrm{miss}}), (i=1,2,3)$ is 
the azimuthal angle between the three leading jets and \met.

The data is compared to the full-detector GEANT4 simulation which is reconstructed 
with the same algorithms as for the data. The Standard Model background processes 
considered in this analysis are QCD (PYTHIA), W/Z+jets (ALPGEN + HERWIG + Jimmy) 
and \ttbar~ (MC@NLO + HERWIG + Jimmy). The PYTHIA QCD predictions 
were compared to a set of ALPGEN QCD samples; the differences were found to be 
well within the experimental uncertainties for the kinematic region explored. 
The QCD and W+jets backgrounds are normalized to the data in the control regions 
defined as \met~ $<$ 40 GeV and \mt~ $<$ 40 GeV for the QCD background and 
30 $<$ \met~ $<$ 50 and 40 $<$ \mt~ $<$ 80 GeV for the W+jets background 
in the one-lepton channel. The QCD background is normalized to the data 
in a control region with at least two jets with \pt~ $>$ 30 GeV and 
leading jet \pt~ $>$ 70GeV in the no-lepton channel.
As an example, the SU4 supersymmetric point (ISAJET+ HERWIG) 
is also shown in the plots with its cross section multiplied by 10; 
SU4 is a low-mass benchmark point close to the Tevatron limits and 
is defined as $m_0$ = 200 GeV, $m_{1/2}$ = 160 GeV, $A_0$ = -400 GeV, 
tan\ensuremath{\beta} = 10 and \ensuremath{\mu} > 0.

The following most important sources of systematic uncertainties are considered: 
the uncertainty on the jet energy scale (which varies from 7-10\% as 
a function of the jet \pt~ and \ensuremath{\eta}), 
the uncertainty on the W+jets and QCD normalizations (50\%), 
the uncertainty on the Z+jets normalization (60\%) and 
the uncertainty on the luminosity (11\%).
\subsection{Results}
\subsubsection{No-lepton channel}
\begin{figure}
\includegraphics[width=.33\textwidth]{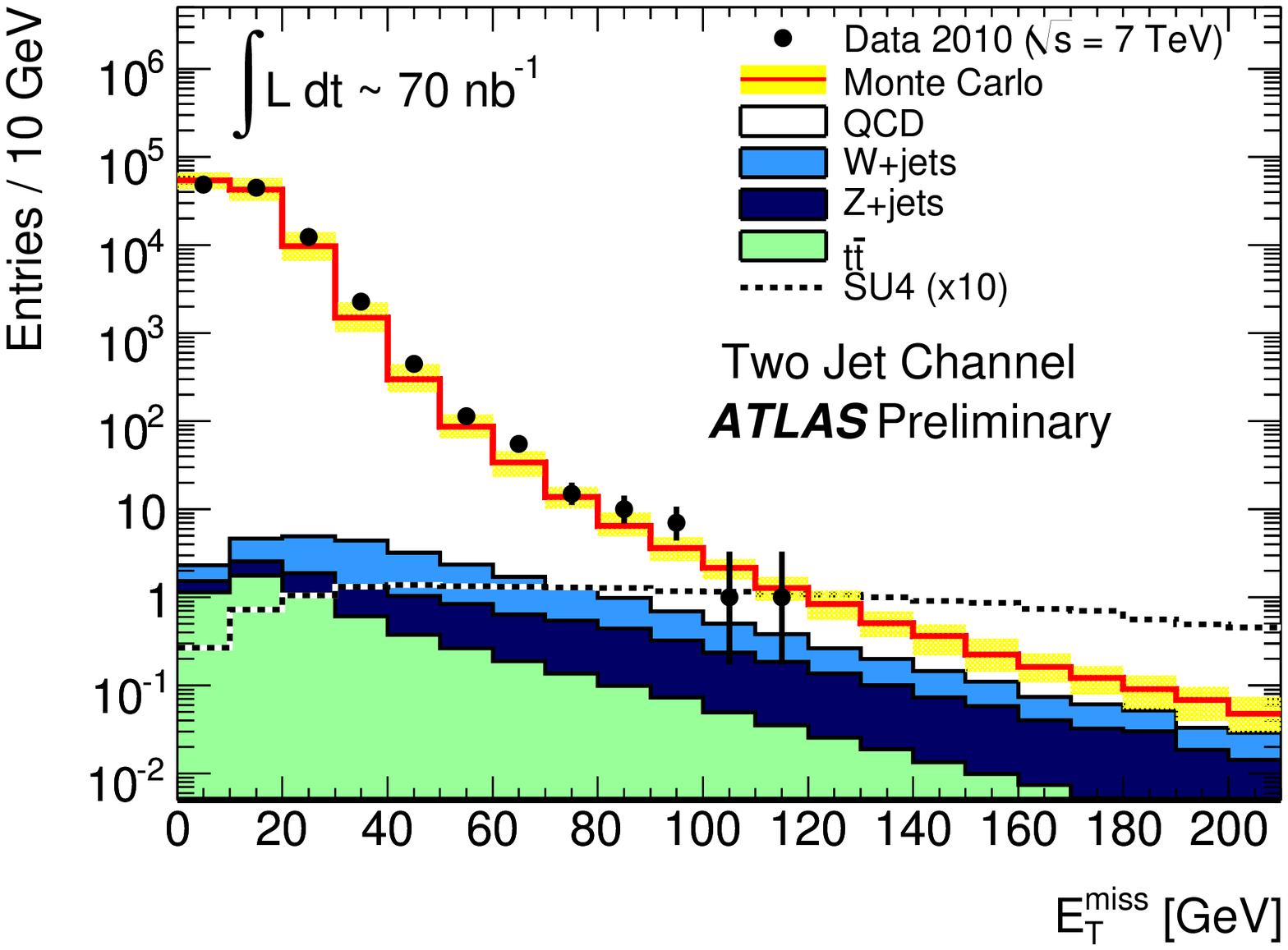}
\includegraphics[width=.33\textwidth]{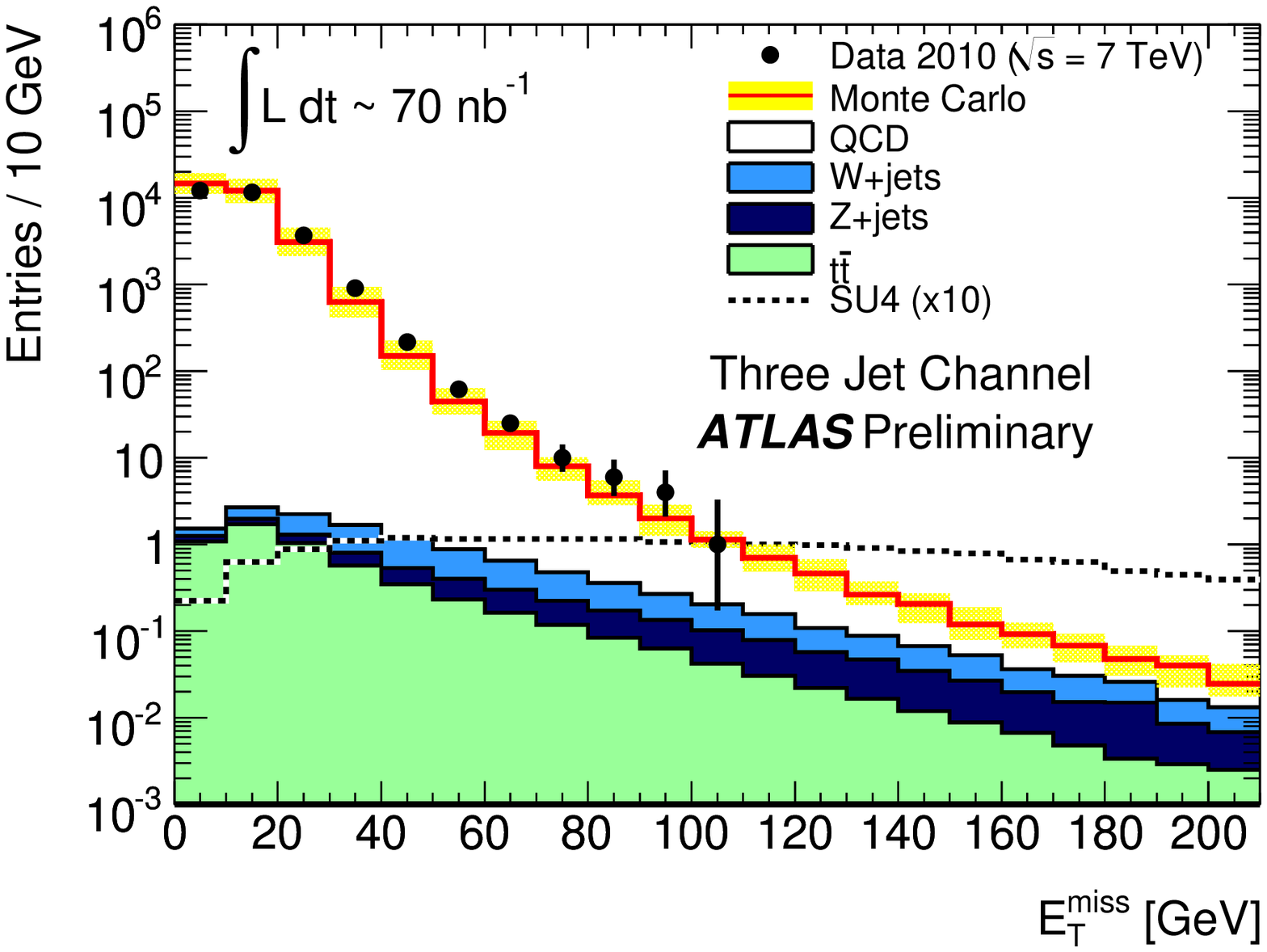}
\includegraphics[width=.33\textwidth]{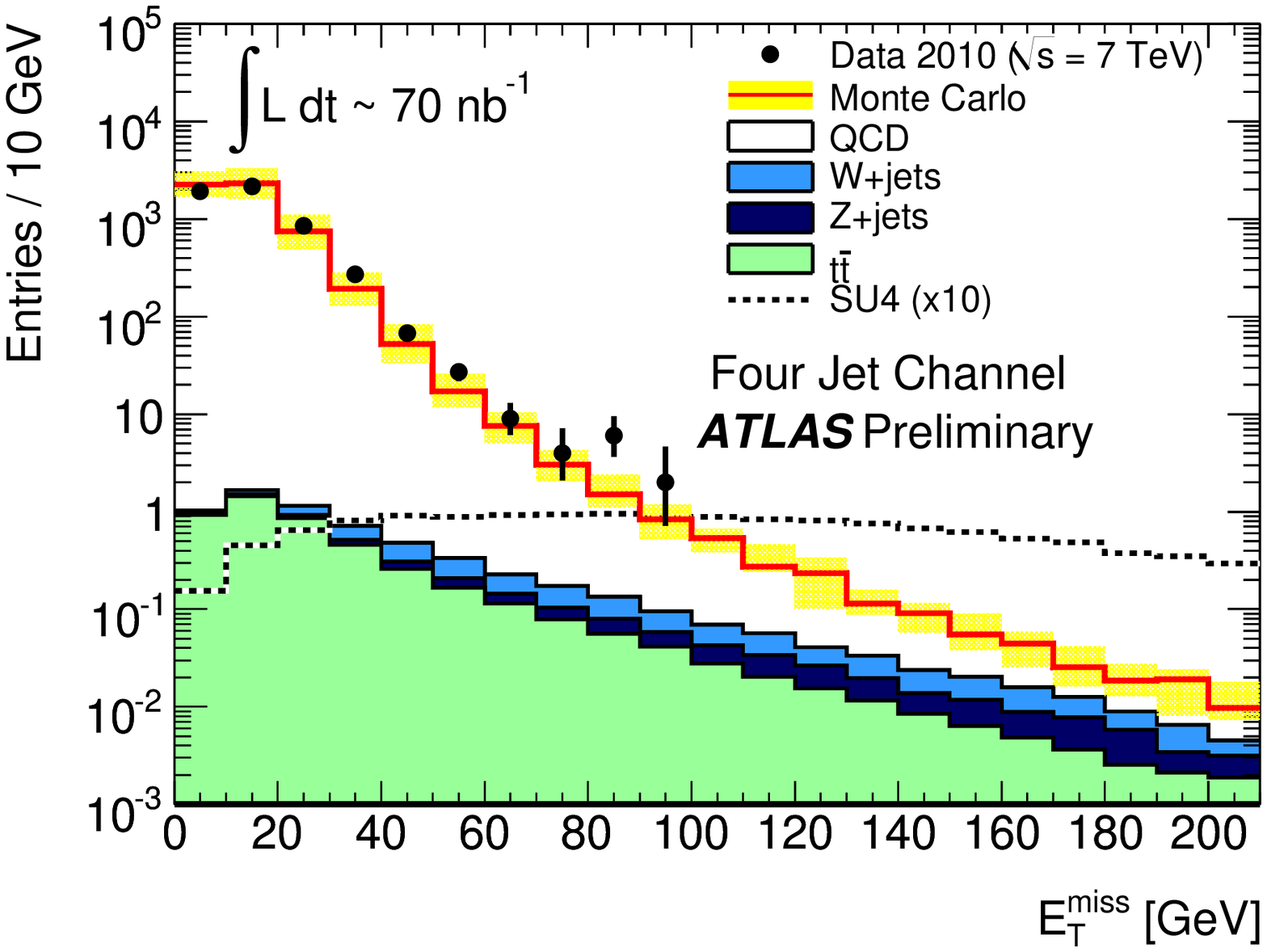}\\
\includegraphics[width=.33\textwidth]{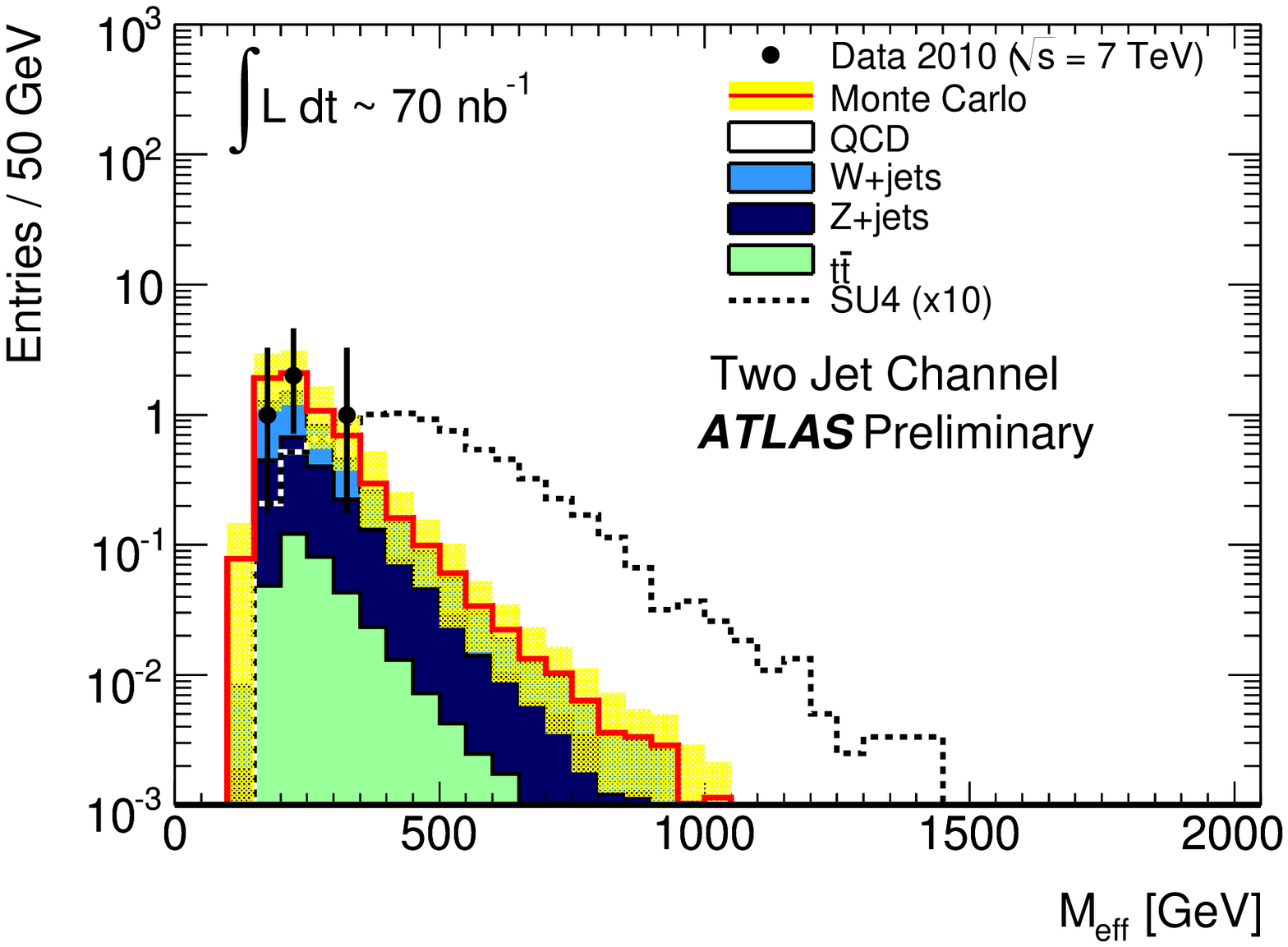}
\includegraphics[width=.33\textwidth]{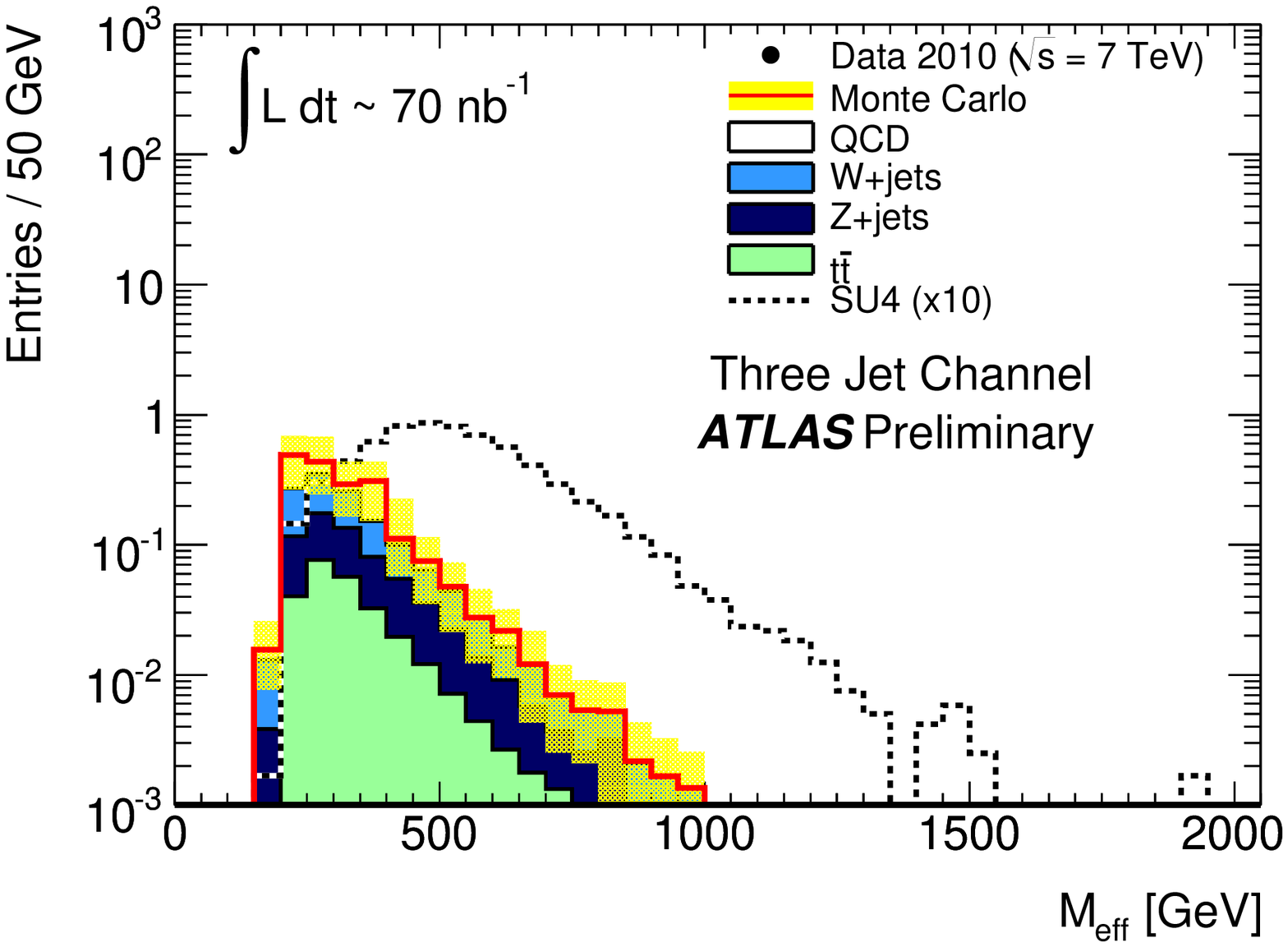}
\includegraphics[width=.33\textwidth]{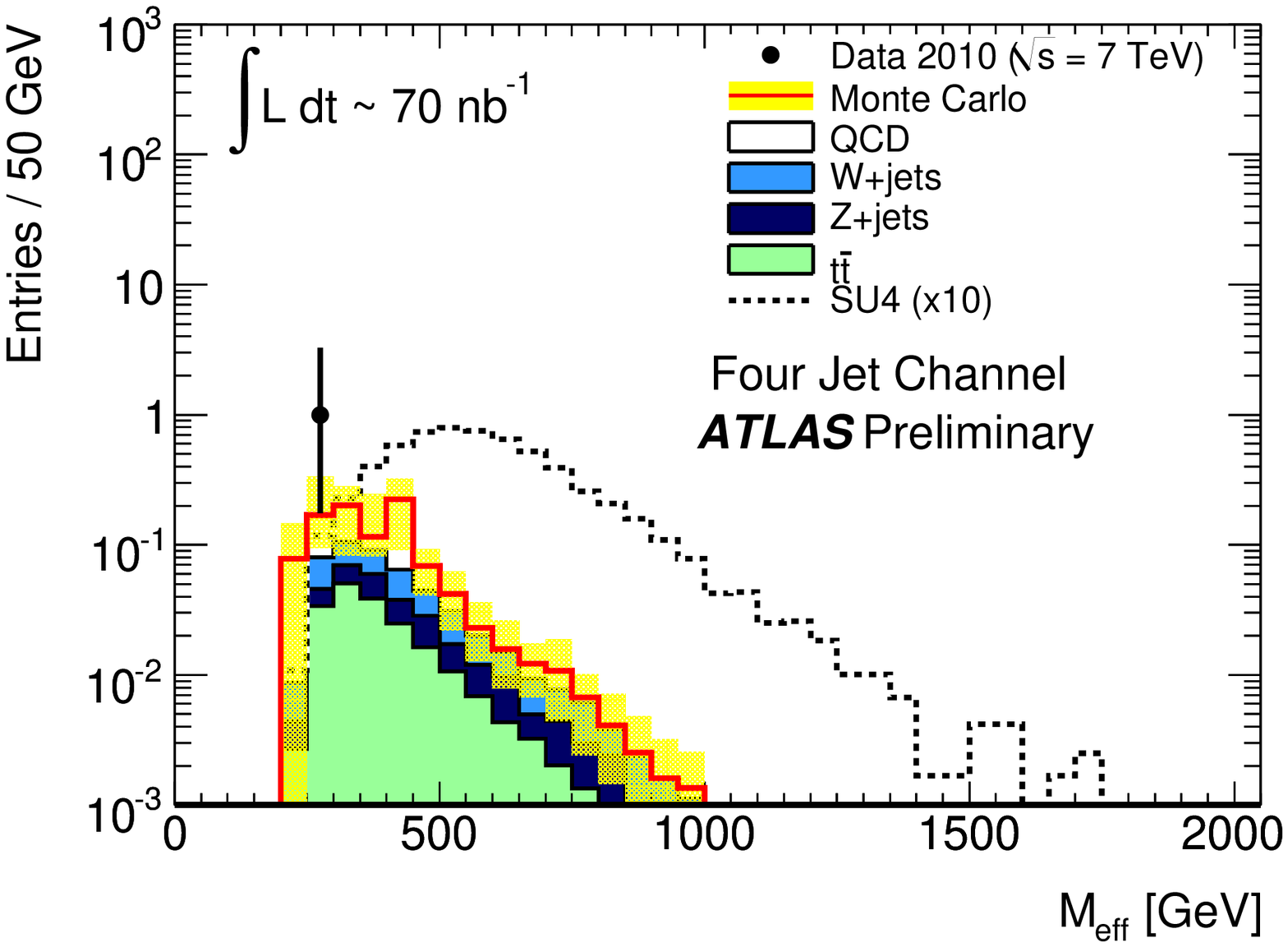}
\caption{Distributions of \met~ (upper) 
  and \meff~ (lower) for events in the two-jet channel (left), 
  three-jet channel (medium) and four-jet channel (right) 
  only with the jet selection cuts (upper) or after 
  final selection cuts (lower) in the no-lepton channel.}
\label{fig_0l}
\end{figure}

Fig.~\ref{fig_0l} shows the \met~ and \meff~ distributions 
for data and the different Standard Model contributions after the 
no lepton event selection in the different jet channels. 
All distributions are reasonably well described
by the Monte Carlo predictions within the systematic uncertainties.
In the signal region, four data events are found, 
which is consistent with the expectation 
of $6.6\pm3.0$ in the two-jet channel, no data events are found, which is
consistent with the expectation of $1.9\pm0.9$ in the three-jet channel 
and one data event is found, which is consistent with the 
expectation of $1.0\pm0.6$ in the four-jet channel 
with $70\pm8$ $nb^{-1}$ of integrated luminosity\cite{SUSY0L}.


\subsubsection{One-lepton channel}
\begin{figure}
\includegraphics[width=.48\textwidth]{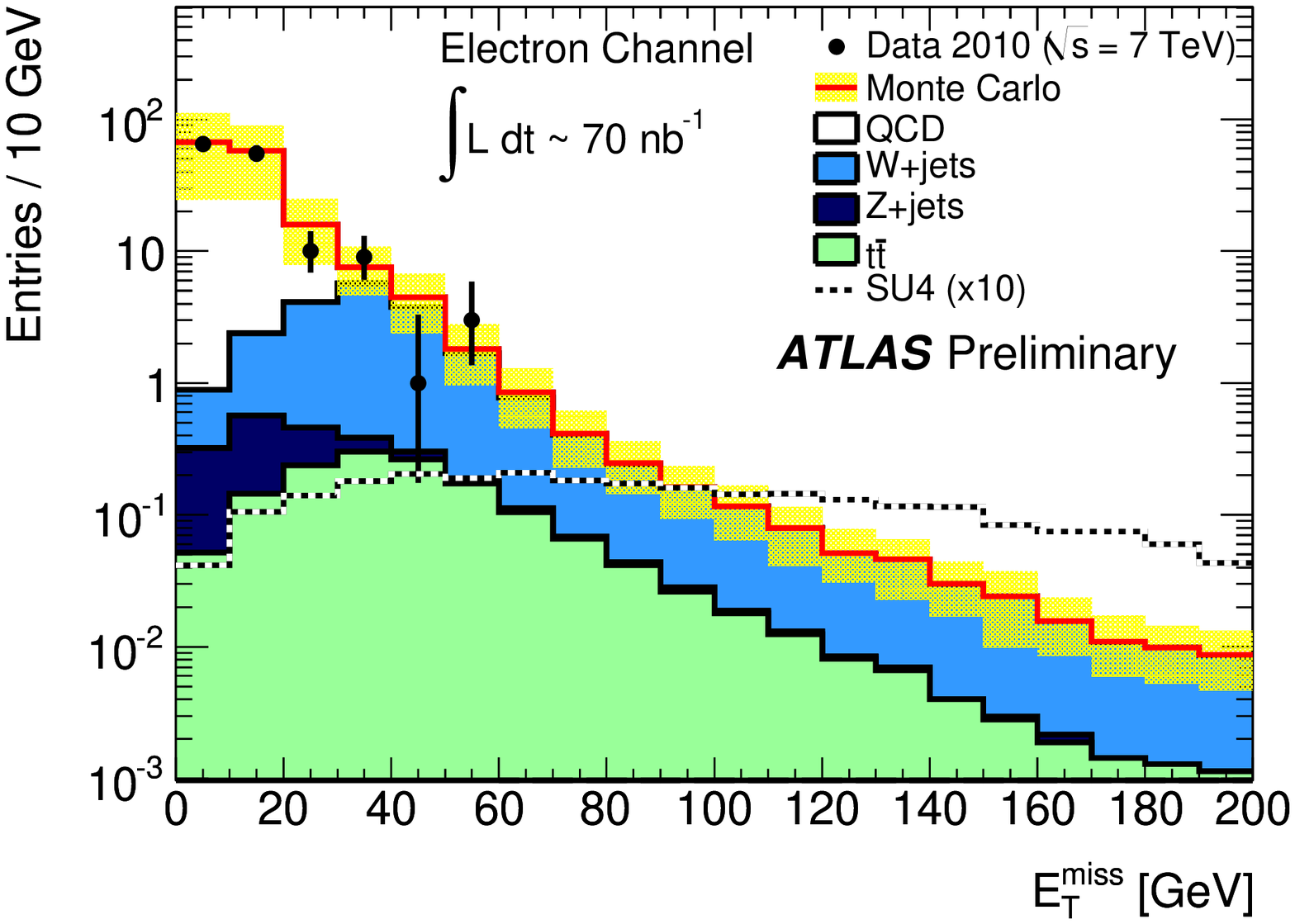}
\includegraphics[width=.48\textwidth]{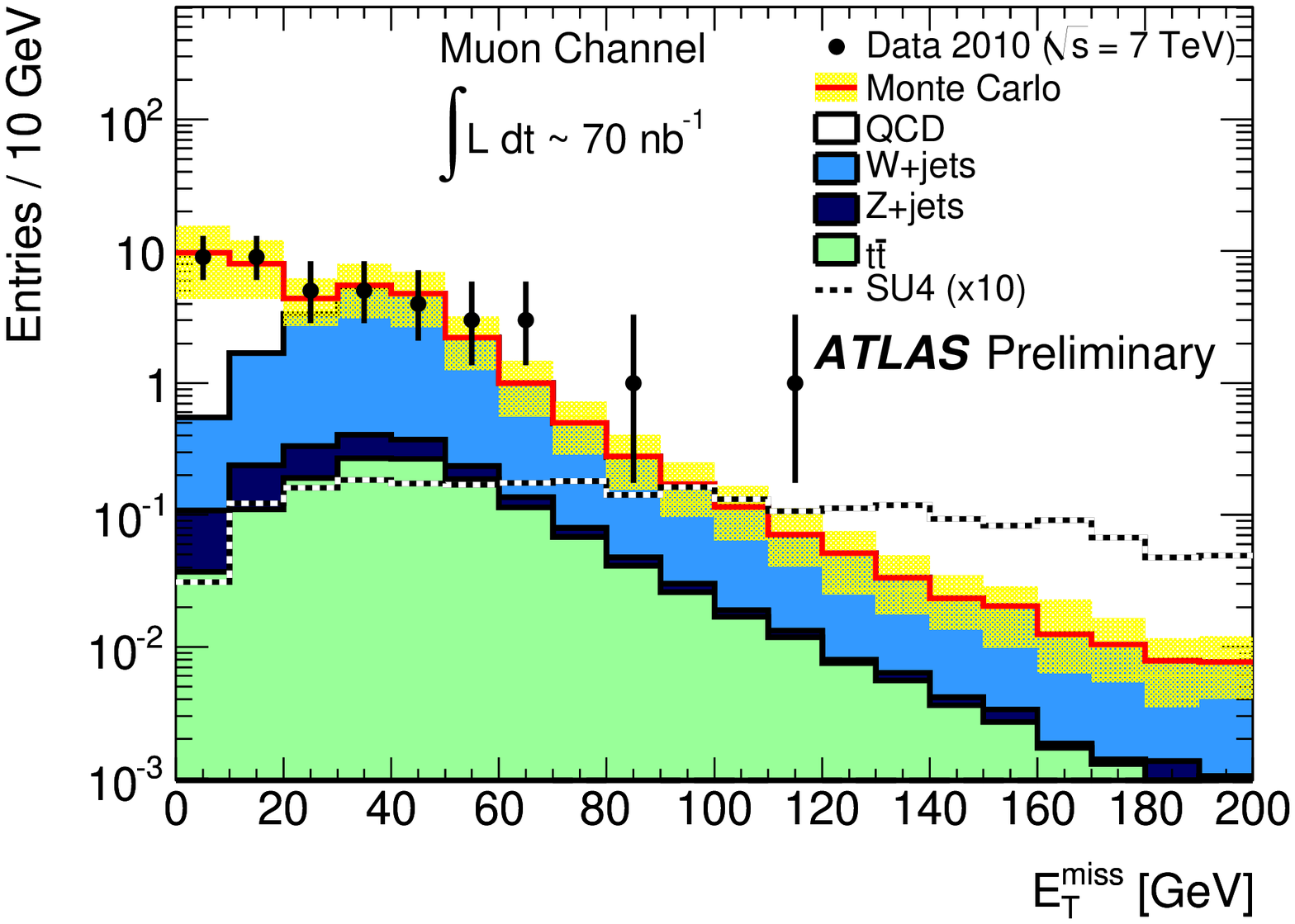}\\
\includegraphics[width=.48\textwidth]{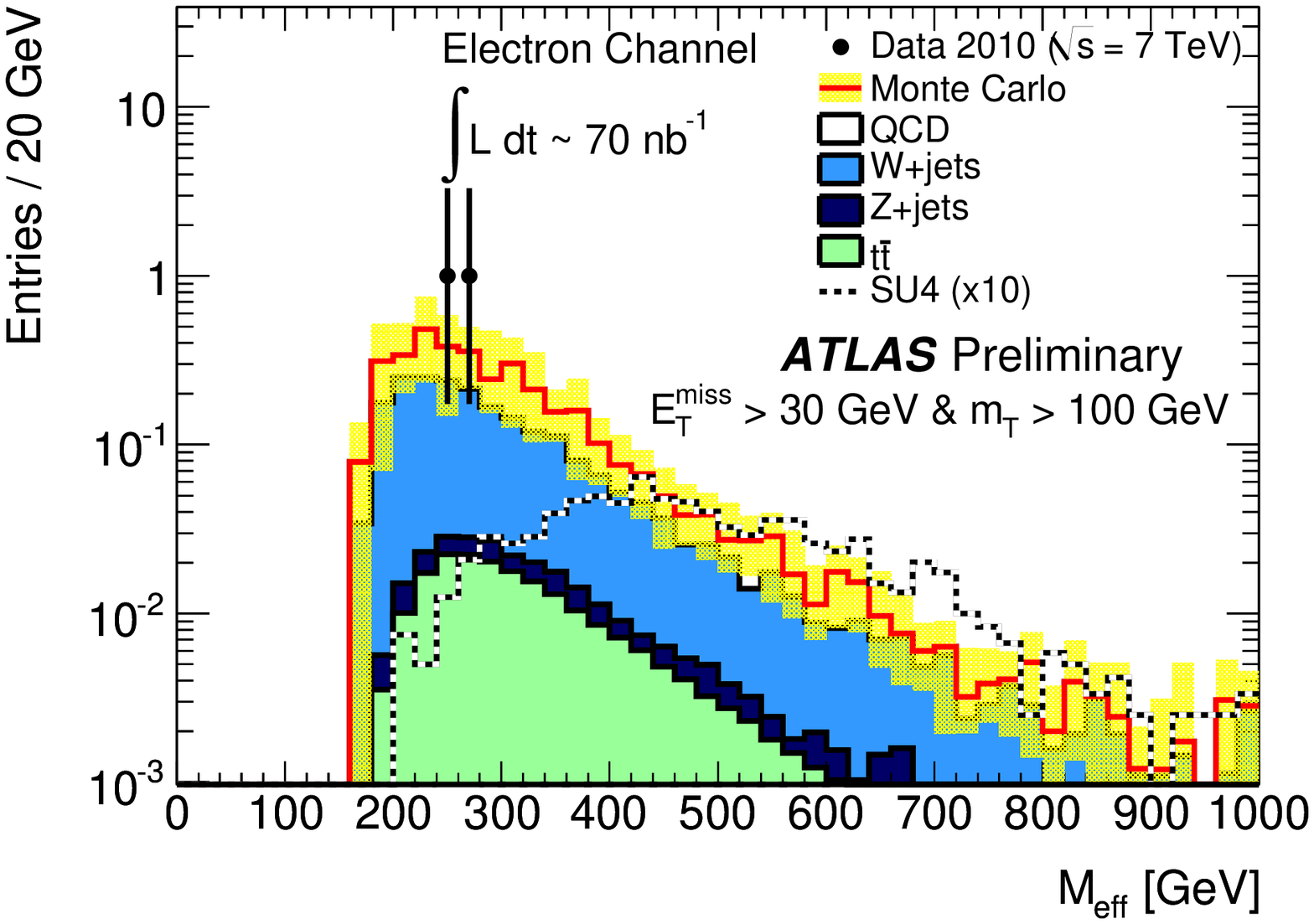}
\includegraphics[width=.48\textwidth]{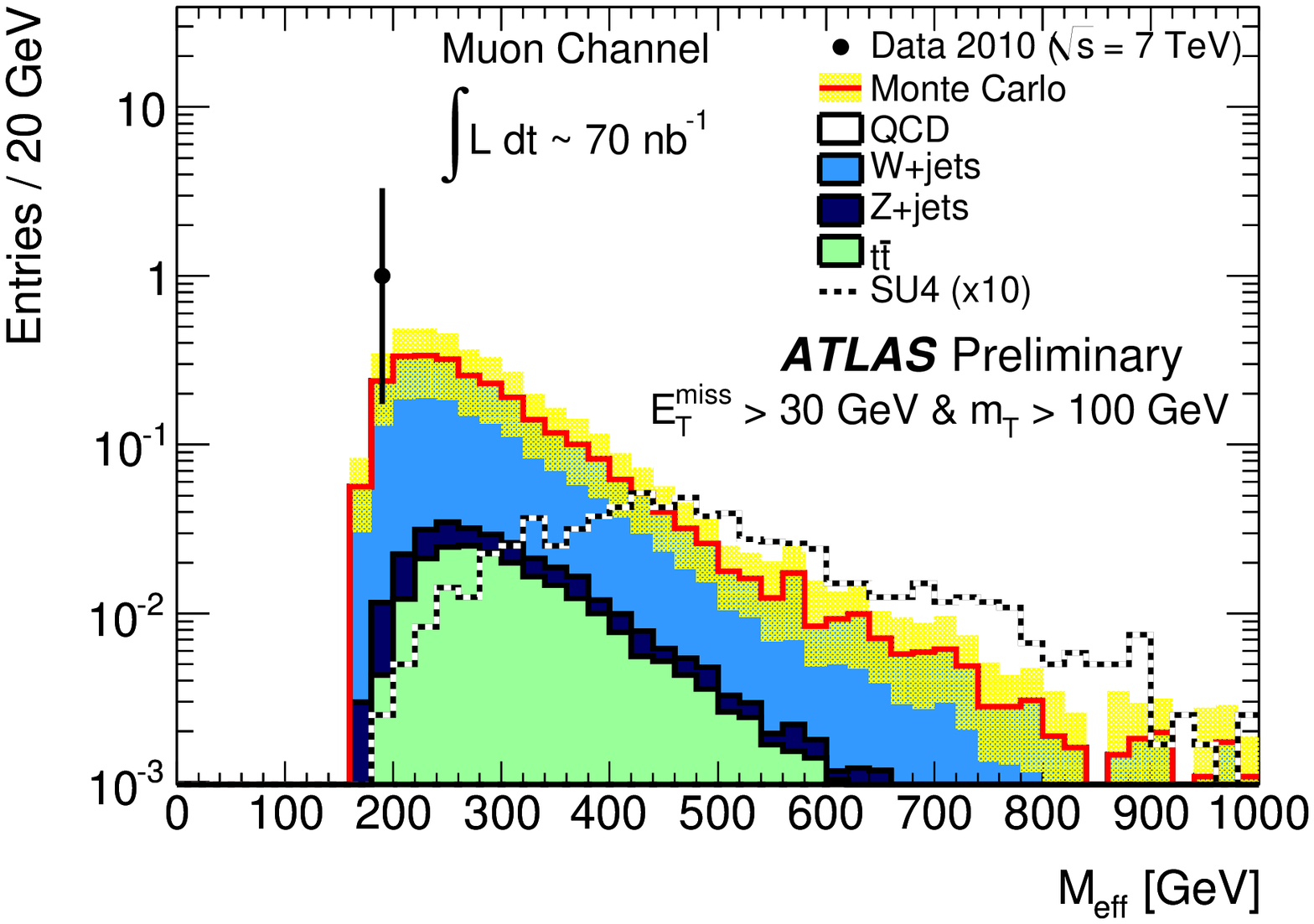}
\caption{Distributions of the missing transverse momentum (\met) for events 
  in the electron channel (upper left) and muon channel (upper right) 
  only with the jet selection cuts requiring and distributions of 
  the effective mass (\meff) for events in the electron channel (left) and muon 
  channel (right) after final selection cuts in the one-lepton channel.}
\label{fig_1l}
\end{figure}

Fig.~\ref{fig_1l} shows the \met~ and \meff~ distributions 
for data and the different Standard Model contributions after the 
electron channel event selection (left) and the muon channel 
event selection (right) in the one-lepton channel. 
All distributions are reasonably well described
by the Monte Carlo predictions within the systematic uncertainties.
Two events remain in the single electron channel and one 
event is found in the single muon channel. The Standard Model expectation 
is $3.6\pm1.6$ events in the electron channel and $2.8\pm1.2$ events 
in the muon channel with $70\pm8$ $nb^{-1}$ of integrated luminosity\cite{SUSY1L}.

\subsubsection{b-jet channel with one lepton}
In the framework of minimal supersymmetry (MSSM), the
production of third generation squarks could be favoured.
Direct pair production of sbottom or stop quarks can lead 
to a final state consisting of a pair of bottom-quark 
jets (b-jets) and significant \met. Leptons might also be 
present\cite{SUSYBJET}.

\begin{figure}
\includegraphics[width=.48\textwidth]{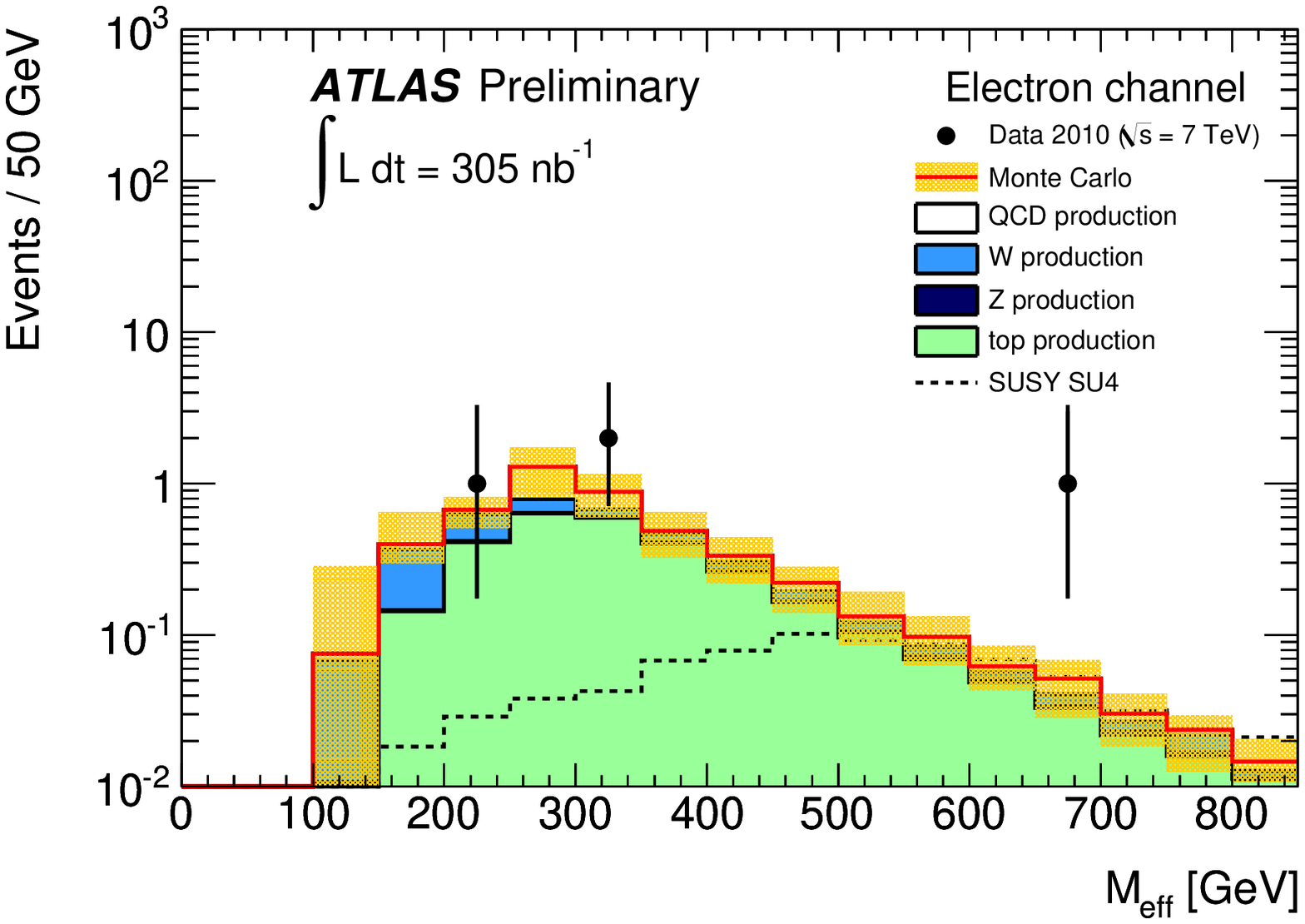}
\includegraphics[width=.48\textwidth]{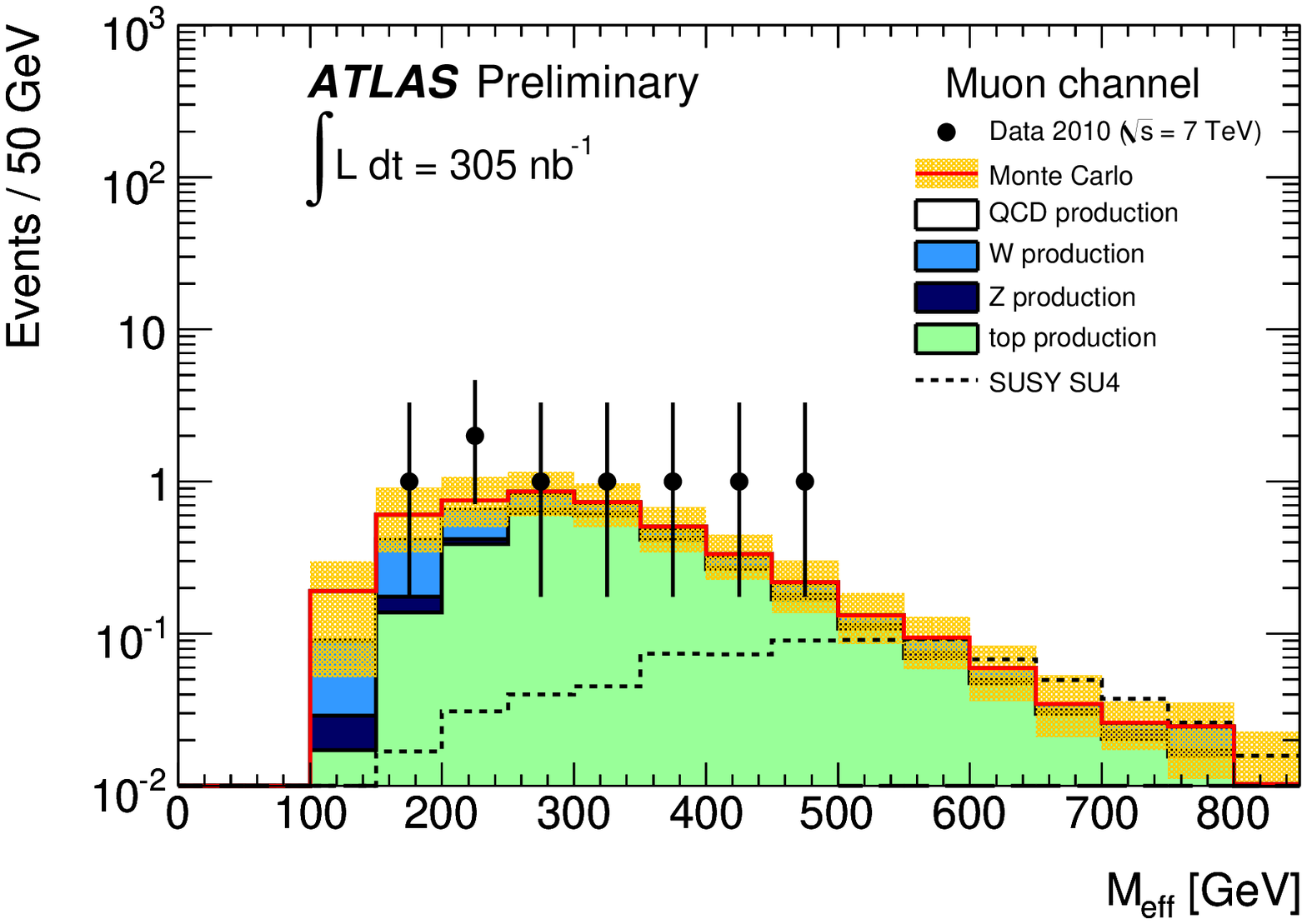}
\caption{Effective Mass (\meff) distributions for data and the different 
  Standard Model contributions after the electron channel (left) and 
  muon channel (right) event selections in the b-jet channel. The uncertainty 
  band includes statistic and systematic uncertainties. 
  The SU4 supersymmetry benchmark point is also shown.}
\label{fig_bjet}
\end{figure}

Fig.~\ref{fig_bjet} shows the \meff~ distributions for data and the different 
Standard Model contributions after the electron channel 
event selection (left) and muon channel event selection (right) in the b-jet channel. 
After applying all selections cuts, 4 events remain for the electron channel, 
with a Standard Model expectation of $4.8^{+1.7}_{-1.5}$, and 8 events remain 
for the muon channel, with a Standard Model expectation of $4.7^{+1.7}_{-1.5}$ 
with $305\pm17$ $nb^{-1}$ of integrated luminosity. 
In both cases, data are in agreement with the Monte Carlo simulation 
within statistical and systematic uncertainties.

\section{Prospects for SUSY discovery}
\begin{figure}
\includegraphics[width=.5\textwidth]{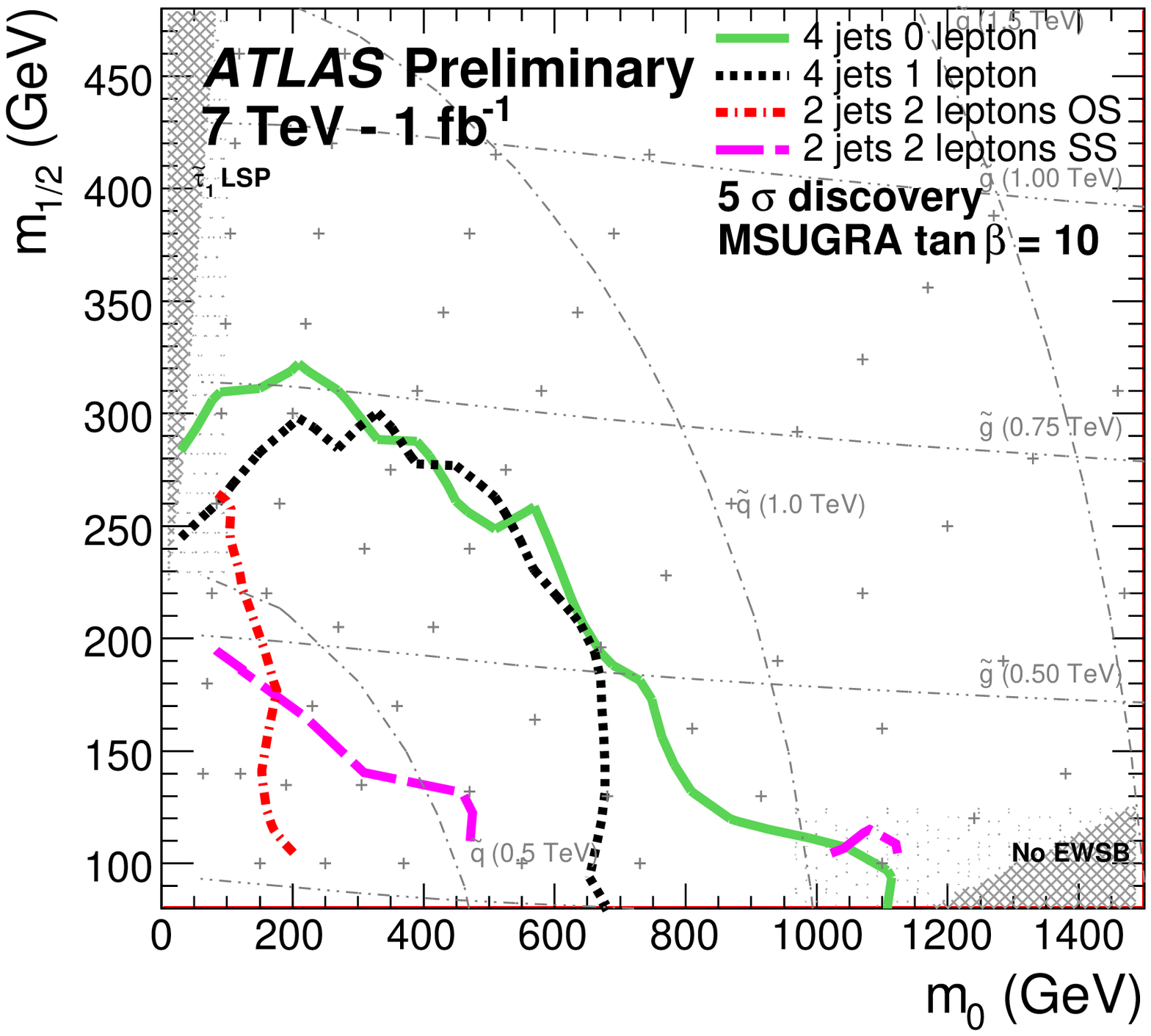}
\includegraphics[width=.43\textwidth]{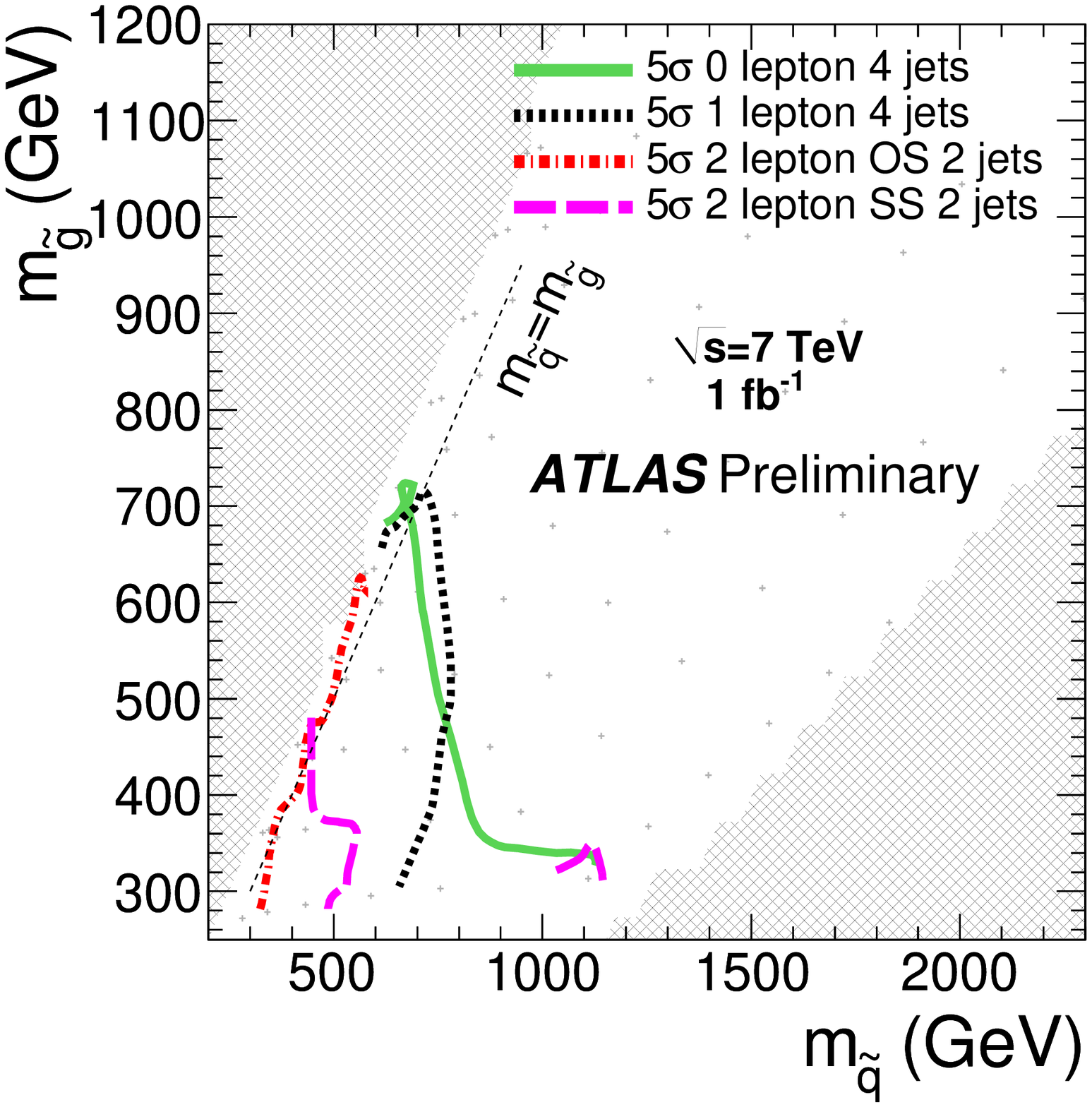}
\caption{5$\sigma$ discovery reach as a function of $m_0$ and $m_{1/2}$ 
masses (left) and of squark/gluino masses (right) for tan$\beta$ = 10 mSUGRA scan 
for channels with 0, 1 and 2 leptons. The integrated luminosity is 1 $fb^{-1}$.}
\label{fig_prospects}
\end{figure}

The search strategy has been applied to grids of models 
in the parameter space of mSUGRA and pMSSM to investigate the discovery reach\cite{SUSYPros}.
Fig.~\ref{fig_prospects} shows the 5$\sigma$ discovery reach as a function of 
$m_0$ and $m_{1/2}$ masses (left) and of squark/gluino masses (right) 
for tan$\beta$ = 10 mSUGRA scan for channels with 0, 1 and 2 leptons for 
a 1 $fb^{-1}$ scenario. The 4-jet 0-lepton and 4-jet 1-lepton channels 
have the largest reach. Signals with squark (first and second generation) 
and gluino masses up to 700 GeV could be discovered in a scenario were 
the gluino mass is similar to the mass of the light squarks.

\section{Conclusion and outlook}
Measured distributions of missing transverse momentum, effective mass 
show agreement with the Standard Model predictions, which 
demonstrates that the ATLAS detector is performing well and that the Monte Carlo 
simulations describe both the underlying physics, and the detector 
response to jets and \met~ within the systematic uncertainties achievable 
thus far. 

The results of the scans indicate that 
ATLAS can discover signals of R-parity conserving SUSY with squark and 
gluino masses up to 700 GeV for L = 1 $fb^{-1}$ in the mSUGRA scenario.

\end{document}